\documentstyle[aps,preprint]{revtex}
\begin{document}

\baselineskip=20pt

\thispagestyle{empty}

\vspace{10mm}

\begin{center}
\vspace{5mm}
{\Large \bf Nuclear Effects on Charmonium Production}

\vspace{1.5cm}

{\sf   
Xiao-Fei Zhang$^{a,b}$, Cong-Feng Qiao$^{a}$, Xiao-Xia Yao$^{a,b}$
and Wei-Qin Chao$^{a,b}$}\\[2mm]

{\small $^{a}$ CCAST (World Laboratory), Beijing, 100080, P.R. China }\\
{\small $^{b}$Institute of High Energy Physics, 
Academia Sinica, P.O.Box 918-4, 100039, P.R.China}\\[10mm]
\end{center}
\begin{abstract}
$J/\psi$ and $\psi'$ production cross sections in fixed-target experiment
is calculated, considering the contributions from both color-singlet 
and color-octet mechanisms.
The results are applied to the investigations of
the $J/\psi$ suppression 
and the $\psi'/\psi$ ratio problems in p-A collisions.
The results agree with the experimental data as the $(c \bar c)-$nucleon 
absorption cross sections $\sigma_{abs}^8\simeq 10mb $ for 
$(c \bar c)_8$ and $\sigma_{abs}^1\simeq 0mb $ for $(c \bar c)_1$.
The model is further used to investigate A-A collisions when comover 
absorption mechanism is also considered. It is found that the observed
experiment data of $J/\psi$ and $\psi'/\psi$ ratio in S-U collisions
and Pb-Pb collisions can not be explained consistently within this
model. The possibility of QGP formation in S-U and Pb-Pb
collisions is also discussed.
\end{abstract}
\vspace{10mm}
PACS number(s): 13.85.Ni,  25.75.Dw, 12.38.Mh

\vspace{5mm}
\newpage

\centerline{\large{\bf  I. Introduction}}

\medskip

Matsui and Satz proposed that a suppression of $J/\psi$ production in
relativistic heavy ion collisions can serve as a clear signature 
for the formation of a new matter phase Quark Gluon 
Plasma (QGP) \cite{Matsui}.
This suppression effect was observed by NA38 collaboration 
later\cite{NA38}. However, it has been found that $J/\psi$ suppression
exists also in p-A collisions where QGP formation is not possible\cite{pa}. 
The successive theoretical researches pointed out 
that $J/\psi$ suppression could also
exist in hadronic matter (HM), even though by  completely different
mechanisms\cite{Capp}. The anomalous $J/\psi$ suppression was
recently reported by the NA50 collaboration\cite{na50}\cite{na50m} 
and there have been 
a number of attempts to explain  it,  such as the onset 
of deconfinement,
hadronic co-mover absorption and the energy 
loss model\cite{blqgp}\cite{comver}\cite{eloss}. 
To understand  the experimental data clearly, the 
formation and absorption mechanism of $J/\psi$ must be 
studied carefully. 

Quarkonium production has traditionally been calculated in the 
color singlet model. However, it has become clear now that the
 color singlet model fails to provide a theoretical and 
phenomenological explanation of all $J/\psi$ production processes
and it is necessary to include the color octet production channel.
In principle, the $J/\psi$ state is described in a Fock state
decomposition
\begin{eqnarray}
\label{Fock}
|J/\psi> &=& O(1)|c\bar c(^3S_1^{(1)})>+O(v)|c\bar c(^3P_J^{(8)}) g>
+O(v^2)|c\bar c(^1S_0^{(8)})g>+ \nonumber \\
&& O(v^2)|c\bar c(^3S_1^{(1,8)})gg>+O(v^2)|c\bar c(^3D_J^{(1,8)})gg>+...,
\end{eqnarray}
where $^{2S+1}L_J^{(1,8)}$ characterizes the quantum state of the $c\bar c$
in color-singlet or octet\cite{nrqcd}, respectively. 
This expression is valid for
the non-relativistic  QCD (NRQCD) framework 
and the coefficient of each
component depends on the relative three-velocity
$|\vec v|$ of the heavy quark. Under the limit of
$|\vec v|\rightarrow 0$, i.e. $c$ and $\bar c$ remain relatively at rest,
Eq.(\ref{Fock}) recovers the expression for color-singlet picture of
$J/\psi$, where $O(1)\equiv 1$.

The color-octet component $(c\bar c)_8$ seems to play  an
important role in interpreting the experimental data by Collider 
Detector at Fermilab (CDF ) \cite{Braaten}, and 
further studies show that it may also have great influence on the quarkonium
production at other collider facilities \cite{a1,a2,a3}. The investigations
\cite{fixed} on quarkonium hadrproduction at fixed target energies find
that the color-octet contribution to the production cross section is 
very important,
and the inclusion of color-octet production channels removes the large 
discrepancies of the total production cross section between 
experimental data and the predictions of the color-singlet
model. Therefore, it has been suggested that the color-octet 
$(c\bar c)_8 $ would also 
manifest itself in heavy ion collisions\cite{satz8}. 
Based on the above discussion, one can accept the following 
physical picture that 
the charmonium production can be divided into two steps.
The first step is the production of a $c \bar c $ pair.
The $c \bar c $ pairs can be either  $(c \bar c)_1$ 
or $(c \bar c)_8$, which are produced 
perturbatively and almost instantaneously, with a formation time 
$\tau_f\simeq (2m_c)^{-1}\simeq 0.07 fm$ in the $c\bar c $ 
rest frame. 
The second step is the formation of a  physical states of  
$J/\psi$, that needs a much longer time.  People believe now that $J/\psi$ 
suppression in hadron matter can be considered
as the pre-resonance absorption.
Satz first proposed the pre-resonance absorption model 
to explain the nuclear collision data and got some
encouraging results\cite{satz8}. However, in their
work the pre-resonance state of charmonia is only in color-octet,
the color-singlet counterpart is believed having a minor influence,
therefore negligible.
In this paper we make a complete leading order calculation of 
charmonium production cross section for $(c \bar c)_1$ and
$(c \bar c)_8$ in concerned energies. The results are used to study 
the charmonium suppression in heavy ion collisions.

As it is known, in hadron-hadron collisions
the initial $(c \bar c)$ pair produced from the parton interaction
are almost point like. It can be in color-singlet or -octet
configurations. The $(c \bar c)$ pair in the color-octet state
may interact with the nucleon
environment much stronger than that in the color-singlet,
which means that the color-octet pair would be dissolved in a much shorter
time than the color-singlet pair.
Based upon this, the absorptions of $(c\bar c)_1 $
and $(c\bar c)_8 $ by nucleon environment should be considered 
differently.
In the framework of this paper, we suppose that the 
$(c\bar c)-$nucleon absorption cross sections $\sigma_{abs}^1$
for $(c\bar c)_1$ and $\sigma_{abs}^8$ for $(c\bar c)_8$ satisfy the condition  $\sigma_{abs}^1\sim 0<<\sigma_{abs}^8$, however the color-singlet part is not completely
negligible. With this assumption the $J/\psi$ suppression data 
 in p-A collision can be explained.
The experimental data of  $\psi'/\psi$ ratio in p-A collisions can
also be explained quite well. 
For nucleus$-$nucleus (A-A) collisions, in addition, the
comover  absorption is investigated as well 
and we find that  the experimental data of  $J/\psi$
suppression and $\psi'/\psi$ ratio can not be described 
consistently in S-U and Pb-Pb collisions.
The possibility for  the production of QGP is also discussed.


Our paper is divided into four parts. In  section II, we describe 
how to introduce the
color-octet scenario borrowed from 
p-p collisions and calculate the charmonium production cross section
through both color-singlet and color-octet mechanisms.
In section III, the pre-resonance 
nuclear absorption model relating to both 
$(c \bar c)_1$ and $(c \bar c)_8$ is described
and used to discuss the $J/\psi$ and $\psi'$ suppression in 
p-A collisions. In section IV, the nucleon and comover absorption 
model for  $J/\psi$
suppression is investigated. In the last section, some further discussions
are given. \\

\centerline{\large{\bf II. Formulation}}

\medskip

According to the NRQCD factorization formalism, the inclusive production rate
of heavy quarkonium $H$ in parton level can be factorized as 

\begin{eqnarray}
\label{21}
\sigma (i j\rightarrow H ) = \sum\limits_{n}\hat{\sigma} (i j \rightarrow
Q\bar{Q}[n] ) <{\cal{O}}^H [n]>.
\end{eqnarray}
Here, $\hat{\sigma} (Q\bar{Q}[n] + X)$ describes the short distance production
of a $Q\bar Q$ pair in the color, spin and angular momentum state n, which
can be calculated perturbatively using Feynman diagram methods. $<{\cal{O}}^H
[n]>$, the vacuum expectation value of a four fermion operator in NRQCD 
\cite{nrqcd},
describes the nonperturbative transition of the $Q\bar{Q}$ pair hadronizing
into the quarkonium state $H$. 
The relative importance of the various contributions
of $n$ in Eq.(\ref{21}) can be estimated by using 
NRQCD velocity scaling rules. 
An important feature of this equation is that $Q\bar Q$ pairs in a 
color-octet state are allowed to contribute to the production 
of a color singlet
quarkonium state $H$ via nonperturbative emission of soft gluons. Accordingly,
the production cross section for a quarkonium state $H$ in the hadron 
process
\begin{eqnarray}
\label{22}
A + B \rightarrow H + X
\end{eqnarray}
can be written as
\begin{eqnarray}
\label{23}
\sigma_H= \sum\limits_{i,j}\int^{1}_{0}dx_1 dx_2 f_{i/A}(x_1)
f_{j/B}(x_2){\sigma} (i j \rightarrow H ),
\end{eqnarray}
where the parton scattering cross section is convoluted with parton 
distribution functions $f_{i/A}$ and $f_{j/B}$, and the sum runs over 
all partons in the colliding hadrons.

At leading twist and at leading order in $\alpha_s$, the color-singlet
$Q\bar Q$ production subprocesses for $^{2S+1}L_J$ state are 
\begin{eqnarray}
\label{24}
g  g \rightarrow ~^1S_0, ~^3P_{0,2},\\
g  g \rightarrow ~^3S_1+g, ~^3P_{1}+g,\\
g  q \rightarrow ~^3P_{1}+q,\\
q \bar q \rightarrow ~^3P_{1}+ g.
\end{eqnarray}
The corresponding formulae of the above precesses can be found in refs.
\cite{fixed} and \cite{report}. 
Because the radiative decays $\chi_{1,2}\rightarrow J/\psi + \gamma$ are 
known to have
a large branching ratios to $J/\psi$  and the feeddown of the
$\psi'$ to $J/\psi$ is also important, their 
contributions should be included in the calculation 
in reproducing the fixed target experimental data 
of the prompt $J/\psi$ production.

It has long been known that the total cross sections of quarkonum production
are rather large in comparison with the fixed-target experiment 
data with respect to the parton-parton fusion predictions\cite{25}.
However, after including the color-octet production mechanism an overall
agreement can be obtained\cite{fixed}.

Under the NRQCD factorization scheme, to calculate the quarkonium production
cross section, one use a double expansion: the perturbative expansion
of the short distance production amplitude in strong coupling constant 
$\alpha_s$ and the expansion of the nonperturbative 
long distance hadronization 
amplitude in typical velocity of heavy quark inside the heavy quarkonium.
At leading order in perturbative theory and up to next-to-leading order in the 
velocity expansion, the subprocesses for leading-twist $J/\psi$ production
through color-octet intermediate states are
\begin{eqnarray}
\label{25}
q  \bar q \rightarrow c \bar{c} [\b{8}, ^3S_1]\rightarrow J/\psi + X,\\
g  g \rightarrow c \bar{c} [\b{8}, ^1S_0]\rightarrow J/\psi + X,\\
g  g \rightarrow c \bar{c} [\b{8}, ^3P_J]\rightarrow J/\psi + X,\\
q  \bar{q }\rightarrow  c \bar{c} [\b{8}, ^3P_J]\rightarrow \chi_J + X 
\rightarrow J/\psi +\gamma + X.
\end{eqnarray}
Note that the process $q\bar{q}\rightarrow c\bar{c}[\b{8},
^3P_{J}]\rightarrow J/\psi(\psi') + g$ is of higher order in
$v^2$ since the lowest-order nonperturbative transition if forbidden by
charge conjugation, and the amplitude $A(gg\rightarrow c\bar{c}[\b{8},
^3S_1])$ vanishes in leading order in $\alpha_s$, which is consistent with
Yang's theorem which forbids a massive $J=1$ vector boson from decaying to
two massless $J=1$ bosons\cite{yang}. In fact, the theorem requires that the
$A(gg\rightarrow c\bar{c}[\b{8},^3S_1])$ vanish to all orders as the
gluons on the massshell.

The cross sections of Eq.(\ref{25}) are proportional to the NRQCD matrix
elements
\begin{eqnarray}
\label{26}
<0|{\cal O}^{J/\psi}_8 (^3S_1)|0>\sim m_c^3 v^7,\\  
<0|{\cal O}^{J/\psi}_8 (^1S_0)|0>\sim m_c^3 v^7,\\  
<0|{\cal O}^{J/\psi}_8 (^3P_J)|0>\sim m_c^3 v^7,\\  
<0|{\cal O}^{\chi_J}_8 (^3S_1)|0>\sim m_c^3 v^5.  
\end{eqnarray}
It is obvious that the above matrix elements are higher order in $v^2$
compared to the leading color-singlet ones, but their corresponding 
short-distance processes are lower order in $\alpha_s$ than that in 
color-singlet processes. This causes the color-octet processes to make
substantially enhancements in reproducing the fixed target experiment data.

For $\psi'$ production the cross section does not receive contributions from
radiative decays of higher charmonium states. $\sigma (\psi')$
differs from the direct $J/\psi$ production cross section 
$\sigma(J/\psi)_{dir}$ only in the  replacement of $J/\psi$ matrix elements
in Eq.(\ref{26}) by $\psi'$ matrix elements.

Before embarking on the computation of cross sections
the parameters used in the computation should be fixed up.
The uncertainties in the theoretical prediction at fixed-target energies
are substantial and it is impossible at present to extract the
non-perturbative universal color-octet matrix elements by fitting the
theoretical predictions to the experiment data.
The value of $<0|{\cal{O}}_8^{J/\psi(\psi')}(^3S_1)|0>$ used here
is obtained by fitting the theoretical predictions to the 
CDF Collaboration data at large $p_T$. In addition, the number of 
independent matrix elements can be reduced by using the spin symmetry
relations up to corrections of order $v^2$
\begin{eqnarray}
\label{41}
<0|{\cal O}^{\chi_J}_1 (^3P_J)|0> = (2J+1)<0|{\cal O}^{\chi_0}_1 (^3P_0)|0>,\\
<0|{\cal O}^{J/\psi}_8 (^3P_J)|0> = (2J+1)<0|{\cal O}^{J/\psi}_8 (^3P_0)|0>,\\
<0|{\cal O}^{\chi_J}_8 (^3S_1)|0> = (2J+1)<0|{\cal O}^{\chi_0}_1 (^3S_1)|0>.  
\end{eqnarray}
Therefore, the matrix elements $<0|{\cal O}^{H}_8 (^1S_0)|0>$ and
$<0|{\cal O}^{H}_8 (^1P_0)|0>$ enter fixed target production of $J/\psi$ and  
$\psi'$ in the combination
\begin{eqnarray}
\label{42}
\Delta_8(H)\equiv <0|{\cal O}^{H}_8 (^1S_0)|0> + \frac{7}{m_Q^2}
<0|{\cal O}^{H}_8 (^3P_0)|0>.
\end{eqnarray}
Up to corrections in $v^2$, the relevant color-singlet production matrix
elements are related to radial wave functions at the origin or their
derivatives, 
\begin{eqnarray}
\label{43}
<0|{\cal O}^{H}_1 (^3S_1)|0> = \frac{9}{2 \pi}|R(0)|^2,~~
<0|{\cal O}^{H}_1 (^3P_0)|0> = \frac{9}{2 \pi}|R'(0)|^2,
\end{eqnarray}
which can be determined from potential model or from quarkonium
leptonic decays.

The values of these parameters, which we use, are\cite{fixed}
\begin{eqnarray}
\label{44}
<0|{\cal O}^{J/\psi}_1 (^3S_1)|0> = 1.16~GeV^3,~~
<0|{\cal O}^{J/\psi}_8 (^3S_1)|0> = 6.6\times 10^{-3}~GeV^3,\\
<0|{\cal O}^{\psi'}_1 (^3S_1)|0> = 0.76~GeV^3,~~
<0|{\cal O}^{\psi'}_8 (^3S_1)|0> = 4.6\times 10^{-3}~GeV^3,\\
<0|{\cal O}^{\chi_0}_1 (^3P_0)|0>/m_c^2 = 4.4\times 10^{-2}~GeV^3,~~
<0|{\cal O}^{\chi_0}_8 (^3S_1)|0> = 3.2\times 10^{-3}~GeV^3, \\
\Delta_8(J/\psi)= 3.0 \times 10^{-2}~GeV^3,~~
\Delta_8(\psi')= 5.2 \times 10^{-3}~GeV^3.
\end{eqnarray}
In the numerical calculation, we use the Gl\"uck-Reya- Vogt (GRV) leading
order (LO )\cite{grv} parameterization for the parton distributions of
the protons. The c quark mass is fixed to be $m_c=1.5$ GeV and the strong
coupling is evaluated at the scale $\mu = 2 m_c$, that is $\alpha_s
\approx 0.26$. The results of the integrated cross sections for color
singlet and  color octet channels at several different energies are listed
in Table I. We must admit that the results are far from precision because
of some important contributions may be precluded as claimed by
authors\cite{fixed}\cite{tang}. e.g., the effects of higher twist, the
beyond leading order contributions in $\alpha_s$, as well
as the kinematic, etc.. \\

\centerline{\large{\bf III. Pre-resonance Absorption in p-A Collisions}}

\medskip

The conventional survival probability for a $J/\psi$ produced 
in a p-A collision is given by:
\begin{eqnarray}
\label{abs}
S_A&=&{1\over A}{\sigma_{pA}\over \sigma_{pp}}\nonumber\\
&=&\int d^2b dz \rho_A(b, z)
exp\biggl\{-(A-1)\int_z^\infty d z'\rho_A(b,z')\sigma_{abs}\biggr\}\nonumber\\
&=&exp(-L_A \rho_0\sigma_{abs}),
\end{eqnarray}
where $\sigma_{pp}$ and $\sigma_{pA}$ are the $J/\psi$ production 
cross section in proton-proton 
 collisions and proton-nucleus collisions, respectively,
$\rho_A$ is the nuclear density distribution. $\sigma_{abs}$
is the absorption cross section. L is the  effective length of the 
$J/\psi$ trajectory. It can be derived as
\begin{eqnarray}
L&=&{3\over 4}{A-1\over A}r_0 A^{1/3}, \ \  \ \ \  \ for \ heavy\ nucleus\nonumber\\
&=&{1\over 2}{A-1\over A}r_0 A^{1/3}{r_0^2\over {r'}_0^2},
 \  \   \ for \ light \ nucleus,
\end{eqnarray}
where $\rho_0=0.14 fm^{-3}$ and $r_0=1.2 fm$, $r'_0=1.05fm $.

As $c \bar c$ pairs are produced almost instantaneously and the  
formation of the  physical states 
$J/\psi$ or $\psi'$ need a much longer time, people now believe that 
$J/\psi$ and $\psi'$ 
suppression in p-A can be considered
as an absorption of pre-resonance $c\bar c$ pairs. 
As discussed 
in former section, there are both $(c \bar c)_1 $ and $(c \bar c)_8 $
pairs. The color-octet can interact with gluons much more strongly than
the color-singlet $(c\bar c)_1$,
and therefore would dissolve much faster into $D$ and 
$\bar D$ than $(c\bar c)_1$. Thus their absorption cross section 
are different. Considering these facts,
we rewrite Eq.(\ref{abs}) as
\begin{eqnarray}
\label{absos}
S_A&=&{1\over A}{\sigma_{pA}\over \sigma_{pp}}\nonumber\\
&=&f_1\int d^2b dz \rho_A(b, z)
exp\biggl\{-(A-1)\int_z^\infty d z'\rho_A(b,z')
\sigma^1_{abs}\biggr\}\nonumber\\
&+&f_8\int d^2b dz \rho_A(b, z)
exp\biggl\{-(A-1)\int_z^\infty dz'\rho_A(b,z')\sigma^8_{abs}\biggr\}, 
\end{eqnarray}
where $f_1, f_8$ are relative fractions of $(c \bar c)_1$
and $(c \bar c)_8$. $\sigma^1_{abs}, \sigma^8_{abs} $
are the absorption cross sections for $(c\bar c)_1$-nucleon     
and $(c\bar c)_8$-nucleon, correspondingly.     

In Eq.(\ref{absos}) there are two parameters, $\sigma_{abs}^1$ 
and $\sigma_{abs}^8$, which is different from 
Satz's model\cite{satz8}.
As ${(c\bar c)}_1$ produced is almost point like, $\sigma_{abs}^1$ 
is very small. Thus we  take $\sigma_{abs}^1=0$ and the 
value of $\sigma_{abs}^8$ is considered as an open parameter and 
determined such as to get the best agreement with the data.
In Fig.1 we see that with $\sigma_{abs}^8=10mb$  we get quite 
good agreement with the p-A data. 

Next we turn to discuss the ratio $\psi'/\psi$ in p-A collisions.
As $J/\psi$ and $\psi'$ suppression in p-A can be considered
as an absorption of pre-resonance $c\bar c$ pairs, there is no difference
for $J/\psi$ and $\psi'$ in p-A collisions. Then  
the $\psi'/\psi$ ratio in p-A collisions can be expressed as
\begin{eqnarray}
\label{pas}
R_A&=&{B(\psi'\to \mu^+\mu^-)\sigma_{p-A\to\psi'}\over
B(J/\psi\to \mu^+\mu^-)\sigma_{p-A\to J/\psi}}\nonumber\\
&=&{B(\psi'\to \mu^+\mu^-)[\sigma'_1exp(-L_A \rho_0\sigma^1_{abs})
+\sigma'_8exp(-L_A \rho_0\sigma^8_{abs})]
\over B(J/\psi\to \mu^+\mu^-)[\sigma_1exp(-L_A \rho_0\sigma^1_{abs})
+\sigma_8exp(-L_A \rho_0\sigma^8_{abs})]}, 
\end{eqnarray}
where B is the corresponding branch ratio, 
$\sigma_1$, $\sigma_8$ are the production 
cross section of color singlet and color octet for $J/\psi$ in p-p 
collisions, respectively.  $\sigma'_1$, $\sigma'_8$ are the production 
cross section of color singlet and color octet for $\psi'$ 
in p-p collisions, respectively.

Using the same parameters as those in Fig. 1,
the result of Eq. (\ref{pas}) is shown in Fig. 2, where one can see that
the ratios $\psi'/\psi$ obtained in our model, being almost independent
of the c.m.s. energy, agree  with the experimental data quite well.
The results show that including both the contribution of color singlet and color octet 
does not introduce any unusual A dependence. \\

\centerline{\large{\bf IV. Comover Absorption in A-A Collisions}}

\medskip

In A-A collisions, except for $(c\bar c)-$nucleon absorption,
charmonium may also suffer interaction with secondaries
that happen to travel along with them. The $J/\psi$ survival 
probability   due to absorption with comover hadrons is
\begin{equation}
\label{co1}
S^{co}=exp\{-\int d \tau \sigma_{co}n_{co}(\tau,b)\},
\end{equation}
where $ \sigma_{co} $ is the $J/\psi$-comover  absorption
cross section, $n_{co}(\tau,b) $ is the density of the 
comovers at time $\tau$ and impact parameter b. The relative 
velocity between $J/\psi$ and the comover is included in the 
definition  of the absorption cross section.
 
Integrating over time $\tau$, assuming that the comovers undergo 
an isentropic longitudinal expansion, 
Eq.(\ref{co1})
can be expressed as 
\begin{equation}
\label{con}
S^{co}=exp\{-\sigma_{co}n_{0}\tau_0 ln{n_0\over n_f}\},
\end{equation}
or,
\begin{equation}
\label{cot}
S^{co}=exp\{-\sigma_{co}n_{0}\tau_0 ln{\tau_{\psi}\over \tau_0}\},
\end{equation}
where $\tau_0$ is the production time of the comovers and
$\tau_{\psi}$ is the time the comovers and $J/\psi$ stay together.
$n_0$ and $n_f $ are  the initial density and freezout density of 
the comovers separately.
When $\tau_\psi$ is smaller than the life time of the comovers $t_\psi$,
Eq.(\ref{cot}) describes the comover survival probability, 
otherwise  Eq.(\ref{con}) works. For $\psi'$, there are similar equations
except that $\sigma_{co}$ is replaced by $\sigma'_{co}$ which is 
the $\psi'$-comover  absorption cross section.
As the mass of  $\psi'$ is much closer to the $D \bar D$ threshold, only a 50MeV
excitation is needed to break up a $\psi'$, while for $J/\psi$, nearly 650MeV
is needed to be above the $D \bar D$ threshold. Thus
$\sigma'_{co}$ should be  much larger than $\sigma_{co}$.

Considering the  effects of both the  $(c\bar c)-$nucleon absorption
and $J/\psi$-comover absorption, the $J/\psi$ survival 
probability in  A-B collisions is 
\begin{equation}
\label{scp}
S=S^{co}\times S^{nuc},
\end{equation} 
where $S^{nuc}$ is the  $J/\psi$ survival 
probability in  A-B collision due to the $(c\bar c)-$nucleon absorption. It
is similar with Eq.(\ref{absos}) and can be expressed as
\begin{equation}
S^{nuc}=f_1 exp(-(L_A+L_B)\rho_0\sigma_{abs}^1)
+f_8 exp(-(L_A+L_B)\rho_0\sigma_{abs}^8),
\end{equation}
where  $L_A $  and $L_B$ are  the  effective length of the 
$J/\psi$ trajectory in A and  B nucleus correspondingly.
  
From the above equations, one can see that there are some parameters
in the comover model. In Eq.(\ref{con}), the parameters are:
$n_0$, $\sigma_{co}, \tau_0,  n_f$.
In Eq.(\ref{cot}), the parameters are:
$n_0$, $\sigma_{co}, \tau_0,  \tau_\psi$.
We first discuss the data at different $E_T$-bins and analyze 
the $E_T$ dependence of the parameters.
In this paper we wish to adjust  these parameters  consistently
for S-U and Pb-Pb collisions.
The comover density is taken to be proportional to the energy density 
at a certain space-time point, which can be expressed as the density of the 
transverse energy according to Bjorken's assumption\cite{bjo}. Therefore,
the  comover density could be expressed as 
\begin{equation}
\label{n0}
n_0 \sim {E_T\over \Delta V} \sim {E_T\over S(b) \Delta y\tau_0},
\end{equation} 
where $\Delta V$ is the corresponding volume.
$S(b)$ is the  overlapping  area of the two nuclei. $\Delta y $ is the 
corresponding rapidity  windows in the 
central rapidity region. $\Delta y=2.4 $ for S-U collisions and 
$\Delta y=1.2 $ for Pb-Pb collisions.

Considering Eq.(\ref{n0}) the $E_T$ dependence of the comover density 
in S-U and Pb-Pb collisions can be described at the same time based on
collision geometry. 
Now we   use Eq.(\ref{scp}) and Eq.(\ref{con})
to fit the experimental data in S-U and Pb-Pb collisions.
The parameters $\sigma^1_{abs}$ and  $\sigma^8_{abs}$
used in considering the $(c \bar c)$-nucleon absorption in A-B collisions
is taken 
to be the same as those obtained in fitting the data in  p-A collisions. 
We
treat the comover density in  the first $E_T$ bin of S-U collisions, 
$n_0^{1}$,
as an open parameter, adjust $n_0^1, n_f, \sigma_{co} $ to fit 
the $J/\psi$ suppression data in S-U collision,  then choose  
$\sigma'_{co} $ to get  the best fit for $\psi'/\psi$ ratio data
in S-U and Pb-Pb collisions. The results for S-U and Pb-Pb collisions 
are shown in Fig.3-4, with   parameters    
$n_0^1=0.2fm^{-3},  \sigma_{co}=3mb,
\sigma'_{co}=23mb$,  and $n_f=0.1fm^{-3}$.

If  $\tau_\psi<t_\psi$, one should use  Eq.(\ref{cot}) to describe
the comover absorption. It is reasonable to choose 
$\tau_\psi$ proportional to the square root of the
transverse overlapping   area of the two nuclei
\begin{equation}
\label{tf}
\tau_\psi\sim \sqrt{S(b)}.
\end{equation}
Using Eqs.(\ref{n0}), (\ref{tf}), (\ref{cot}) and (\ref{scp}), 
with the parameters 
$n_0^1=0.2fm^{-3},  \sigma_{co}=3mb, \sigma'_{co}=13mb$, and 
$\tau_\psi^1=6.5fm$,
which is the 
$\tau_\psi$ for   the first $E_T$ bin in S-U collisions, the 
results are  shown in Fig. 5 and Fig. 6.

Fig.3-6 show that neither of the two comover absorption
expressions of   Eq.(\ref{con})
and Eq.(\ref{cot}) can  explain the data of $\psi'$
and $J/\psi$ suppression in S-U and Pb-Pb collisions consistently.
Fig. 3(a) shows that the $J/\psi$ suppression in S-U collisions 
could be described very well,  based on the above  chosen parameters 
for Eq.(\ref{con}), however Fig.3(b) shows that the $\psi'/\psi$ 
ratio data in S-U collisions could not be 
fitted using the same set of parameters. The data show an 
anomalous $\psi'$ suppression from the second 
$E_T$ bin. Furthermore, 
Fig. 4(a) shows that the comover absorption  which explains the $J/\psi$
suppression in S-U collisions 
can  not explain the $J/\psi$
suppression in Pb-Pb collision,  where an  anomalous
suppression exists from the second $E_T$ bin.  This seems to show that 
the anomalous $\psi'$
suppression begins already in S-U collisions at an energy density 
much lower than the corresponding density for anomalous $J/\psi$ 
suppression in Pb-Pb collisions. 
This may reflect   the fact that 
the dissociation temperature for $J/\psi$ in QGP is higher than 
the dissociation temperature of $\psi'$\cite{gao}. 
From Fig.4(b) one can find that the same  comover absorption  
can explain the data of $\psi'/\psi$ ratio in Pb-Pb collisions. 
This could  be explained as that the anomalous $J/\psi$ and 
$\psi'$ suppressions are 
canceled in the  $\psi'/\psi$ ratio data. 
The results of using Eq.(\ref{cot}) to include the comover absorption 
 are shown in Fig. 5-6, 
which are  similar to Fig. 3-4.

Now we turn to discuss the case of minimum biased data, where
the result is shown  in a simple and clear way.
With the parameters $n_f=0.1 fm^{-3}, \ \ \tau_0=1 fm $, 
$\sigma_{co}=3mb, \sigma'_{co}=23mb$ which is the same as
that we used in obtaining  Fig.3-4,   taking the average comover density 
 ${\bar n}_{su}=0.28$ in S-U collision and   the average 
comover density ${\bar n}_{pb}=0.4$ in  Pb-Pb collisions, 
the results of using Eq.(\ref{scp}) and Eq.(\ref{con})
for the  minimum biased data are  shown in Fig. 7. 
From it one can see clearly that using the comover absorption
expressions of  Eq.(\ref{con}) the $J/\psi$ suppression data for S-U collisions is fitted based on the above parameters, but one 
can  not explain the data of $\psi'$
suppression in S-U collisions, neither the $J/\psi$ suppression in 
Pb-Pb collision. The good fitting of $\psi'/\psi$ ratio data in Pb-Pb collisions may caused by the same anomalous absorption of $\psi'$ and $J/\psi$.
The results of using  Eq.(\ref{scp}) 
and Eq.(\ref{cot}) is similar to Fig. 7.\\

\centerline{\large{\bf V. Results and Discussions}}

\medskip

In this paper, 
$J/\psi$ and $\psi'$ production cross section is calculated 
considering the contributions of both color-singlet 
and color-octet $c\bar c$ channels.
The pre-resonance absorption model for charmonium 
is extended to consider both the color singlet and color 
octet contribution. Using this model the $J/\psi$
and $\psi'$ suppression in p-A collision are explained very well.
Based on above calculation, the comover absorption  
is discussed  for A-A collisions and
it is found that 
the observed experimental data of $J/\psi$ and 
$\psi'/\psi$ ratio  in S-U collision and Pb-Pb collision
can not be explained consistently by this mechanism.
This indicates that other sources of charmonium suppression 
should be included. 

The situation in explaining the data of strangeness production is 
similar. To explain the  enhanced production of 
strangeness, some kind of 
collective interaction among individual excited nucleon states,
such as  the colour ropes in RQMD\cite{ta1}, the multiquark clusters in 
VENUS\cite{ta2}, firecracker in LUCIAE \cite{ta3}, must be considered.
Some  authors also have reported the 
possibility 
of QGP formation at CERN  SPS from the strangeness puzzle\cite{strang}.
There is no strict way to distinguish   these different kinds of 
collective motions from the formation of  QGP till now.

Now we consider the possibility of QGP formation from our discussion.
In QGP, charmonium breaks up due to color screen. 
As the radius of $\psi'$ is larger than that of $J/\psi$, 
the critical temperature $T'_c$
for $\psi'$ to  be dissociated in QGP should be much lower 
than  the critical
temperature  $T_c$ for $J/\psi$ to  break up\cite{gao}.
So if QGP is formed, $\psi'$ will  begin to  break up earlier
than  $\psi$. 
As has been pointed out above, Fig. 3(b) and Fig.5(b) indicate that 
from the second $E_T$ bin(or the third $E_T$ bin) in S-U collisions 
there is an anomalous 
$\psi'$ suppression, which may be the effect of QGP production.
Fig.4(a) and Fig.6(a) show that    
the anomalous suppression of $J/\psi$  really begins 
after the second $E_T$ bin in Pb-Pb collision
because it needs a higher critical temperature.
If  deconfined phase is attained, when $J/\psi$
begin to be suppressed due to Debye screen, $\psi'$
should already be broken up by Debye screen. These two anomalous
suppressions in QGP  may cancel each other in $\psi'/\psi$ ratio, 
which leads to the 
result that $\psi'/\psi$ ratio in Pb-Pb collisions seems to  be explained by 
the comover absorption, while in fact, neither $J/\psi$ nor
$\psi'$ anomalous suppression could be explained by comover absorption. 
Fig.4(b) and Fig.6(b)
agree with the above  picture of $J/\psi$ and $\psi'$
suppression by QGP. 

Although our results indicate the possibility of production of QGP in S-U
and Pb-Pb collisions,  and to explain enhanced strangeness production  
some collective motion which can not be distinguished from QGP
formation  must also be considered, further study  is still needed 
to distinguish if QGP are really formed.

\vskip 5mm

\centerline{\Large{\bf Acknowledgments}}

\vskip 5mm
This work was supported in part by the National
Natural Science Foundation of China and the 
Hua Run Postdoctoral Science Foundation
of China.

\newpage
\centerline{\bf Figure Captions}

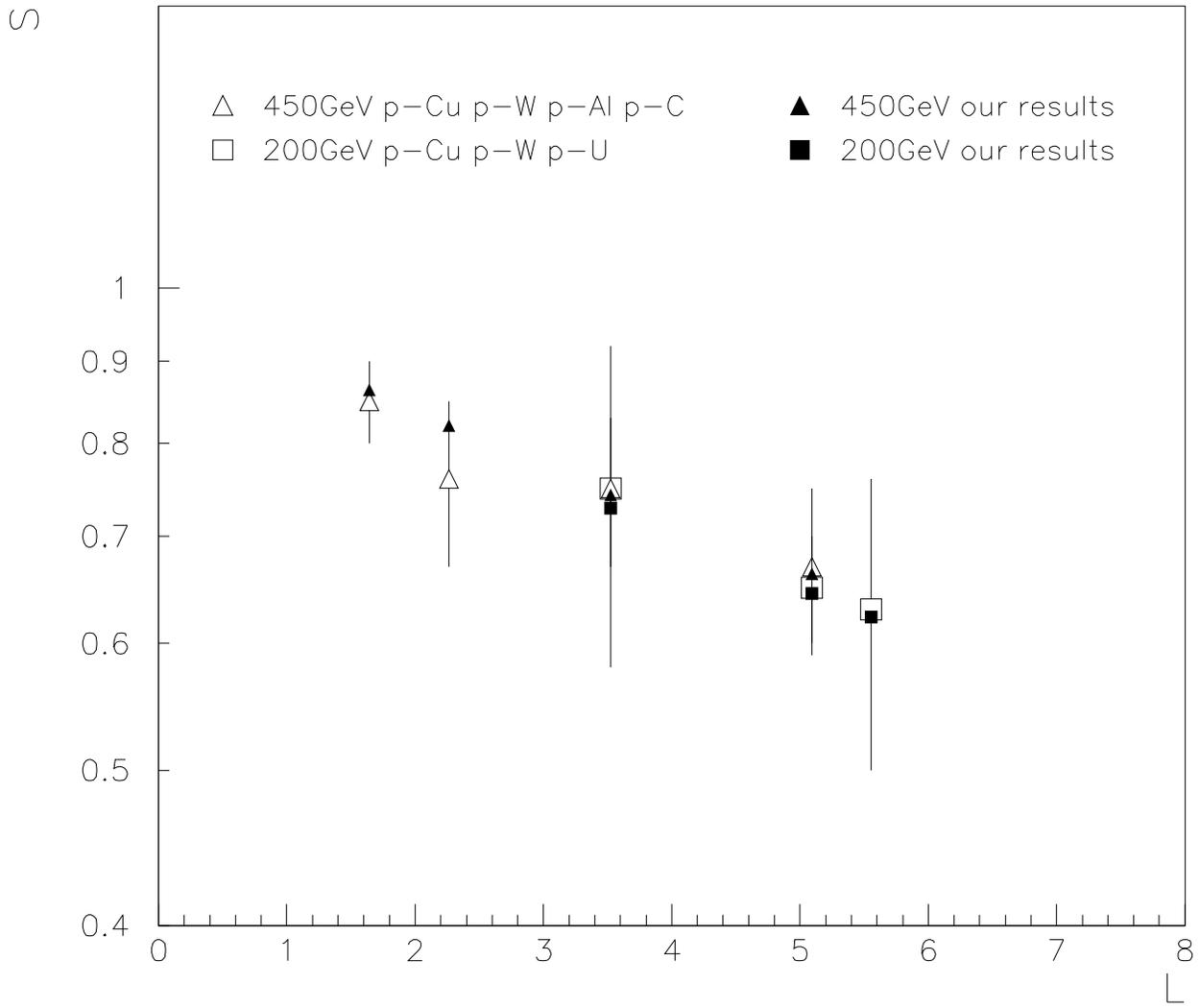
Fig.1: The $J/\psi$ survival probability obtained using 
Eq.(\ref{absos}) with $\sigma_{abs}^8= 10mb $ 
and  $\sigma_{abs}^1=0mb $ is 
compared to the experimental data of p-A 
collisions at different energies.\\

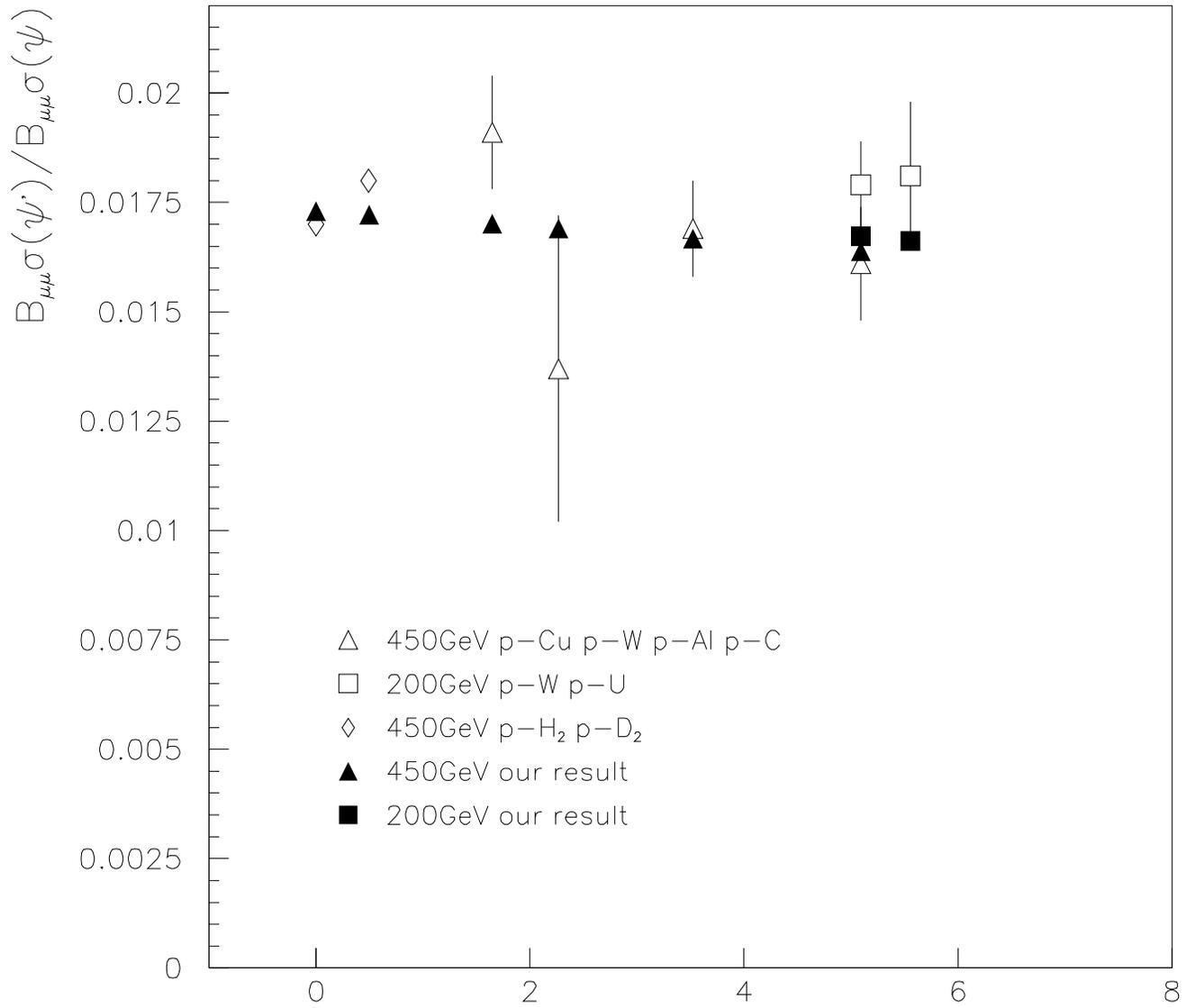
Fig.2: The $\psi'/\psi$ ratio obtained using 
Eq.(\ref{pas}) with $\sigma_{abs}^8= 10mb $ 
and  $\sigma_{abs}^1=0mb $ is 
compared to the experimental data of p-A 
collisions at different energies.\\

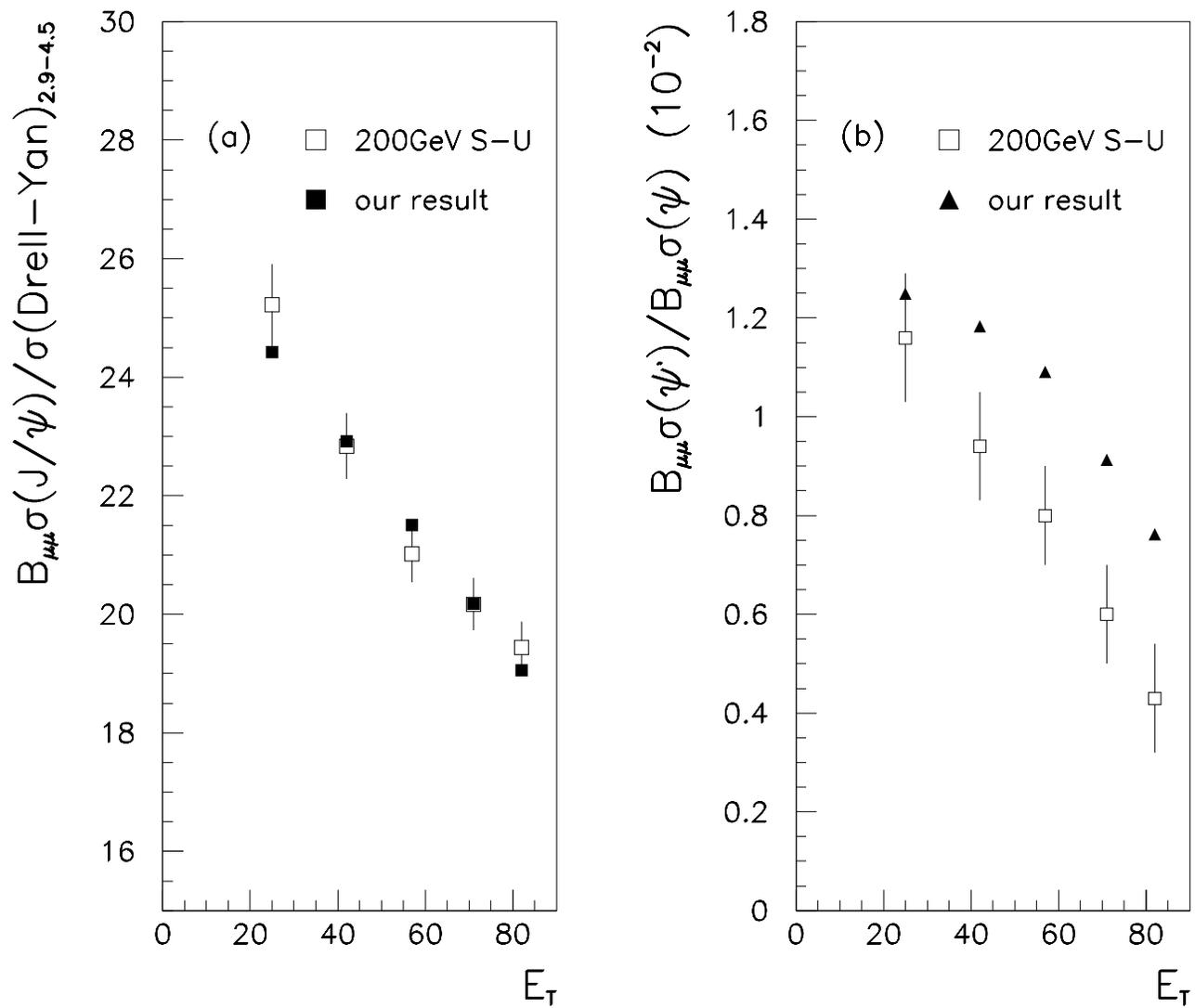
Fig.3: (a) $J/\psi$ over DY ratio versus $E_T$  and (b) 
$B_{\mu\mu}\sigma(\psi')/
B_{\mu\mu}\sigma(\psi)$(b) versus $E_T$ in S-U collision  are 
compared to the results based on Eq.(\ref{scp}) using the comover 
absorption Eq.(\ref{con}). The parameters are 
$n_0^1=0.2fm^{-3},  \sigma_{co}=3mb, \sigma'_{co}=23mb$,  and 
$n_f=0.1fm^{-3}$.\\

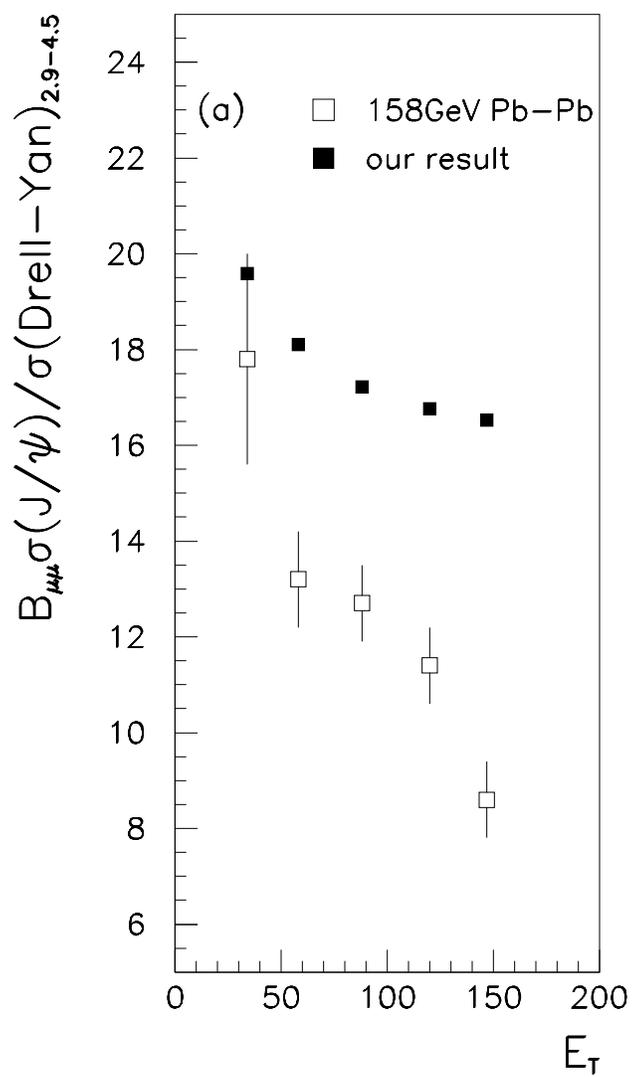
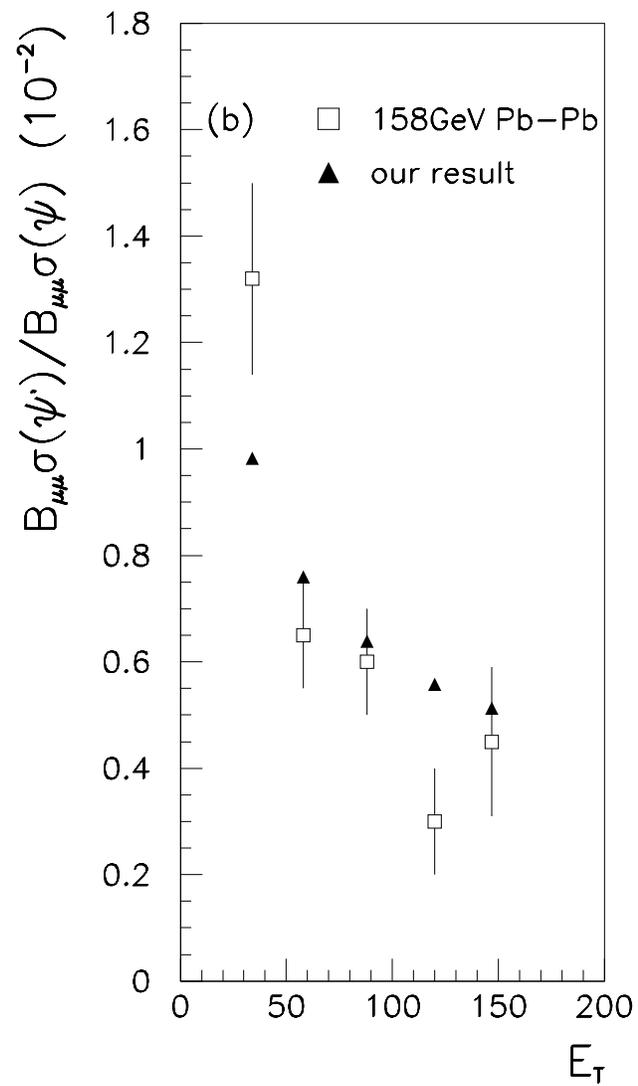
Fig.4: The same as Fig.3 for Pb-Pb collisions.\\

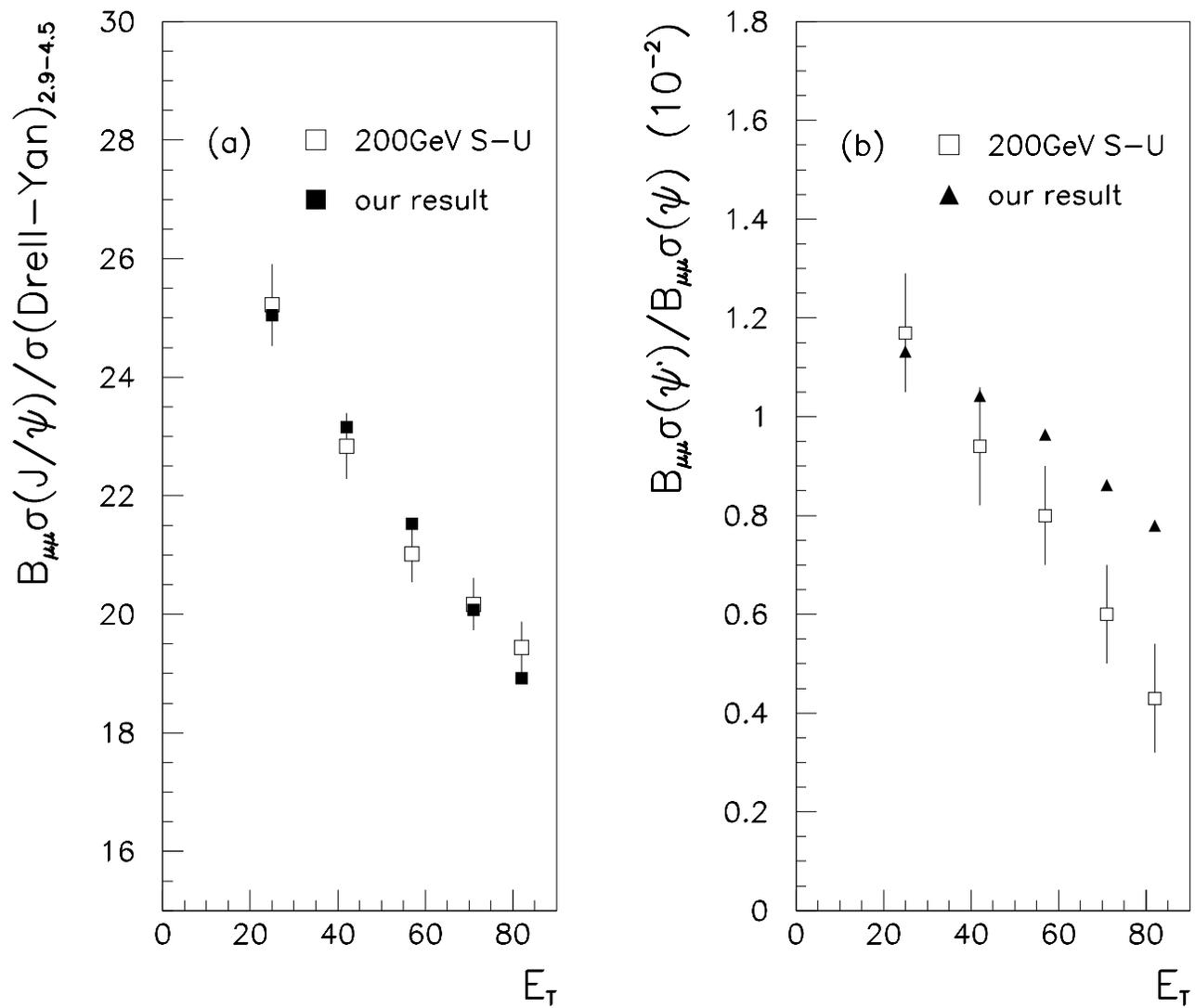
Fig.5: The same as Fig.3 using the comover absorption Eq.(\ref{cot})
with the parameters $n_0^1=0.2fm^{-3}, \sigma_{co}=3mb, \sigma'_{co}=13mb$
and $\tau_\psi^1=6.5fm$.\\

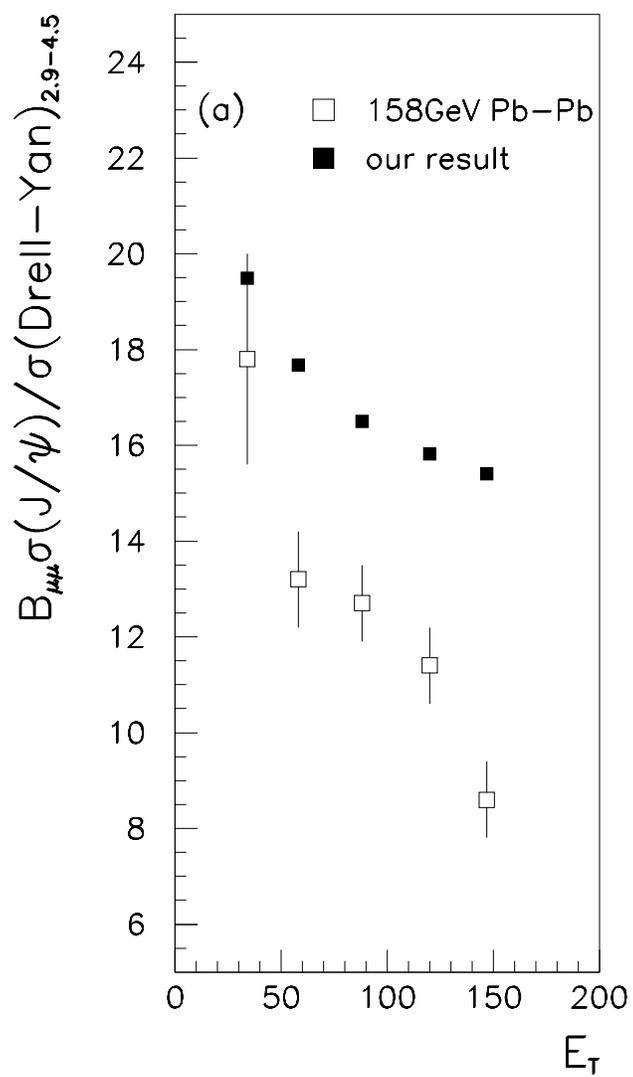
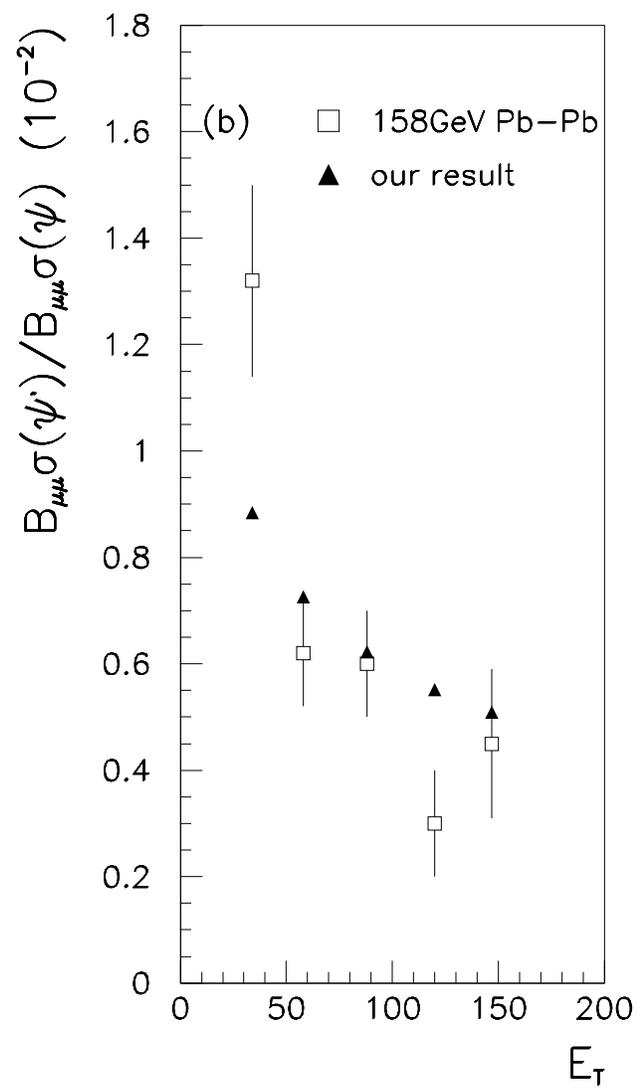
Fig.6:  The same as Fig.5 for Pb-Pb collisions.\\

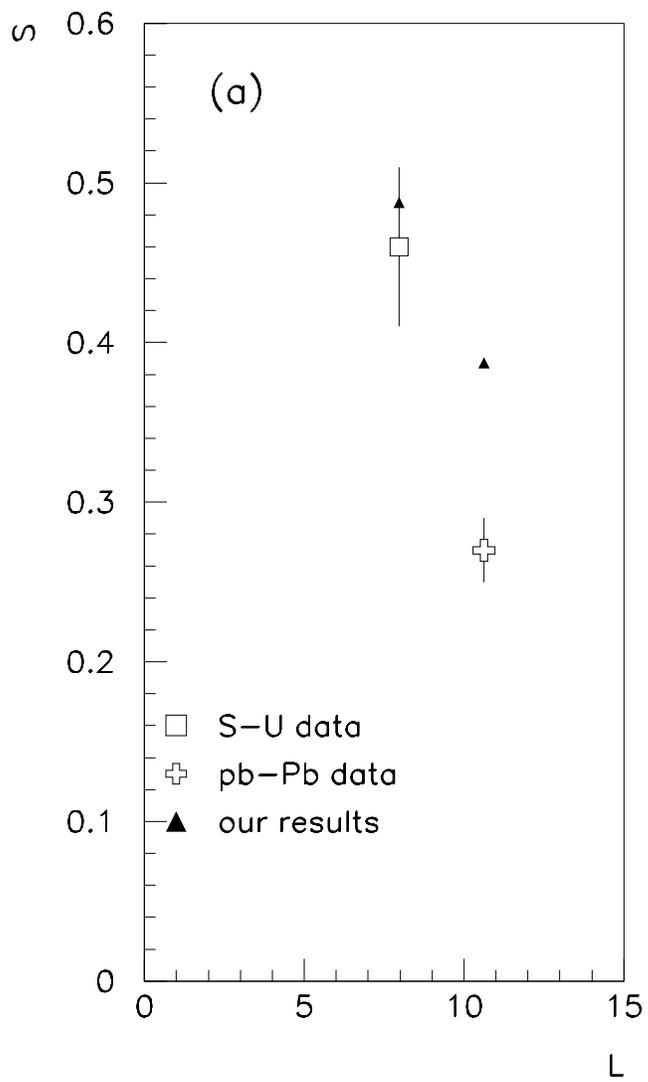
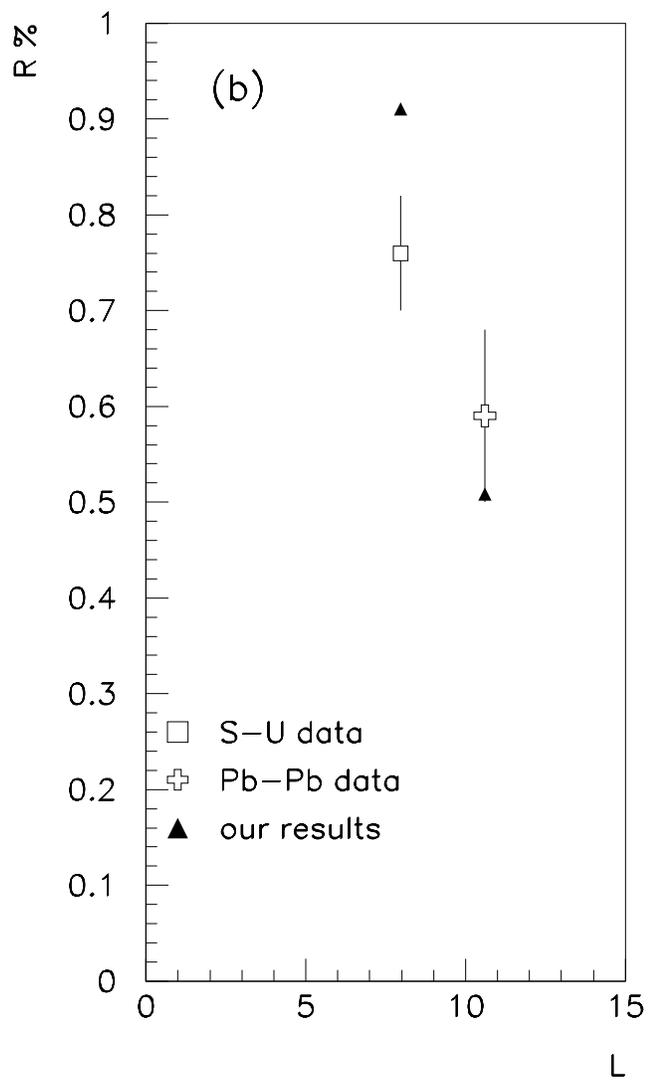
Fig.7: The same as Fig. 3  in the minimum bias case 
with ${\bar n}_{su}=0.28$, ${\bar n}_{pb}=0.4$.\\
    
\vskip 1cm
\centerline {\bf Table Caption}

Table I. The integrated cross sections for color singlet and color octet processes.

\newpage
\begin{center}

Table I

\vskip 1cm

\doublerulesep 2.0pt
\begin{tabular}{cccc} \hline \hline
 & E=450 GeV  & E=200 GeV & E=158 GeV  \\ \hline
$\sigma_1$ & 49.64 nb & 24.56 nb &19.54 nb\\ \hline
$\sigma_8$ & 95.66 nb &  55.48 nb &45.94 nb\\ \hline 
$\sigma'_1$ & 7.76 nb & 3.72 nb &2.86nb\\ \hline
$\sigma'_8$ &20.04 nb &12.11  nb &11.06 nb\\ \hline \hline

\end{tabular}

\end{center}

\vskip 5mm

\end{document}